\documentclass[9pt]{IEEEtran} 

\usepackage{graphicx} 
\usepackage{times}
\usepackage{amsmath,amsfonts}
\usepackage{paralist}
\usepackage{mdwlist}
\usepackage{xspace}
\usepackage{stfloats}
\usepackage{float}
\usepackage{booktabs}
\usepackage{colortbl,color,array}
\usepackage{graphicx}
\usepackage[usenames,dvipsnames]{xcolor}
\usepackage{mathpazo}
\usepackage{listings}
\usepackage{eso-pic}
\usepackage{transparent}
\usepackage{multirow}
\usepackage[font=footnotesize]{caption}
\usepackage{subcaption}
\usepackage{cite}
\usepackage{url}
\usepackage{breakurl}
\usepackage{lipsum}
\newcommand{\subparagraph}{}
\usepackage[compact]{titlesec}
\definecolor{myDarkBlue}{rgb}{0,0.0,0.75} 
\definecolor{myDarkGreen}{rgb}{0,0.45,0.08} 
\definecolor{myantiGreen}{rgb}{0.5,0.0,0.5} 

\begin{document}
%
\title{Precoding by Priority: A UEP Scheme for RaptorQ Codes}

\author{
\IEEEauthorblockN{Keshava M Elliadka \hspace{2 in} Robert Morelos-Zaragoza}\\
\IEEEauthorblockA{Department of Electrical Engineering \hspace{.9 in} Department of Electrical Engineering \\
San Jose State University \hspace{1.7 in} San Jose State University\\
San Jose, CA 95192-0084 USA \hspace{1.4 in} San Jose, CA 95192-0084 USA \\
\hspace{.2 in}Email: \texttt{keshava.elliadka@sjsu.edu} \hspace{.8 in} Email: \texttt{robert.morelos-zaragoza@sjsu.edu}}
}

\maketitle
\begin{abstract} 
Raptor codes are the first class of fountain codes
with linear time encoding and decoding. These codes are recommended in standards such
as Third Generation Partnership Project (3GPP) and digital video 
broadcasting. RaptorQ codes are an extension to Raptor codes, having better coding efficiency
and flexibility. Standard Raptor and RaptorQ codes are systematic with equal error protection
of the data. However, in many applications such as MPEG transmission, there is a need for 
Unequal Error Protection (UEP): namely, some data symbols require higher error correction capabilities compared to others.
We propose an approach that we  call Priority Based Precode Ratio (PBPR)
to achieve UEP for systematic RaptorQ and Raptor codes.
Our UEP assumes that all symbols in a
source block belong to the same importance class.
The UEP is achieved by changing the number of precode symbols depending on   
the priority of the information symbols in the source block. PBPR provides
UEP with the same number of decoding overhead symbols for source blocks with different importance classes.
We demonstrate consistent improvements in the  error correction capability of
higher importance class  compared to the 
lower importance class across the entire range of channel erasure probabilities. We also show that PBPR does not result in a significant increase in
decoding and encoding times compared to the standard implementation.
\end{abstract}

\section{Introduction} 

Fountain codes \cite{MacKay2005} are a class of rateless error control codes
that are suitable for applications (such as internet) in which the channels resemble 
binary erasure channels with a time-varying erasure probability.
The encoder generates potentially unlimited number of encoding
symbols (known as check or repair symbols) and the decoder can decode the source 
symbols from a large enough subset of these encoding symbols. Luby-Transform (LT) codes \cite{Luby2002} are improved 
universal Fountain codes that have reduced encoding and decoding complexities compared to the 
original.  Raptor codes \cite{Shokr06} are an  extension of LT codes, and are the first fountain codes  with linear time encoding
and decoding. They have a two stage coding scheme wherein the LT codes are concatenated 
with a high rate Low Density Parity-Check (LDPC) code---the \textit{precode}---that enables the recovery of those symbols that are not recovered in the LT decoding phase. 
Raptor codes are used in many standard applications like Third Generation Partnership Project (3GPP) \cite{3gpp.26.346} 
Multimedia Broadcast Multicast Services (MBMS), Digital Video Broadcasting-Handheld (DVB-H) \cite{dvbh}, etc. 
RaptorQ codes are the most recent class of fountain codes. They provide higher flexibility, larger source
block size, and better coding efficiency compared to  Raptor codes \cite{rfc6330}.

Standardized Raptor and RaptorQ codes are systematic codes and provide Equal
Error Protection (EEP) to all information symbols. However, in several applications, a portion of the
information symbols require higher error correction compared to others. An
important example is MPEG transmission where I-frames have higher priority
compared to P-frames and B-frames.  This motivates the need for
\textit{Unequal Error Protection (UEP)} \cite{Masnick67}. UEP schemes classify
symbols into multiple groups, and   provide different error correction to
each group.

Existing UEP schemes for the Raptor family of codes can be broadly classified into two categories. 
The first category provides UEP by modifying the selection probability of symbols. 
Such schemes are applicable to customized Raptor codes \cite{RahVF07,YuaA10}.  The second category of UEP schemes use 
expanding and overlapping windows of source symbols.   UEP is achieved by repeating the 
source symbols in appropriate windows \cite{Sejdin09}. \\


\begin{figure*}
\centering
       \includegraphics[width=\columnwidth]{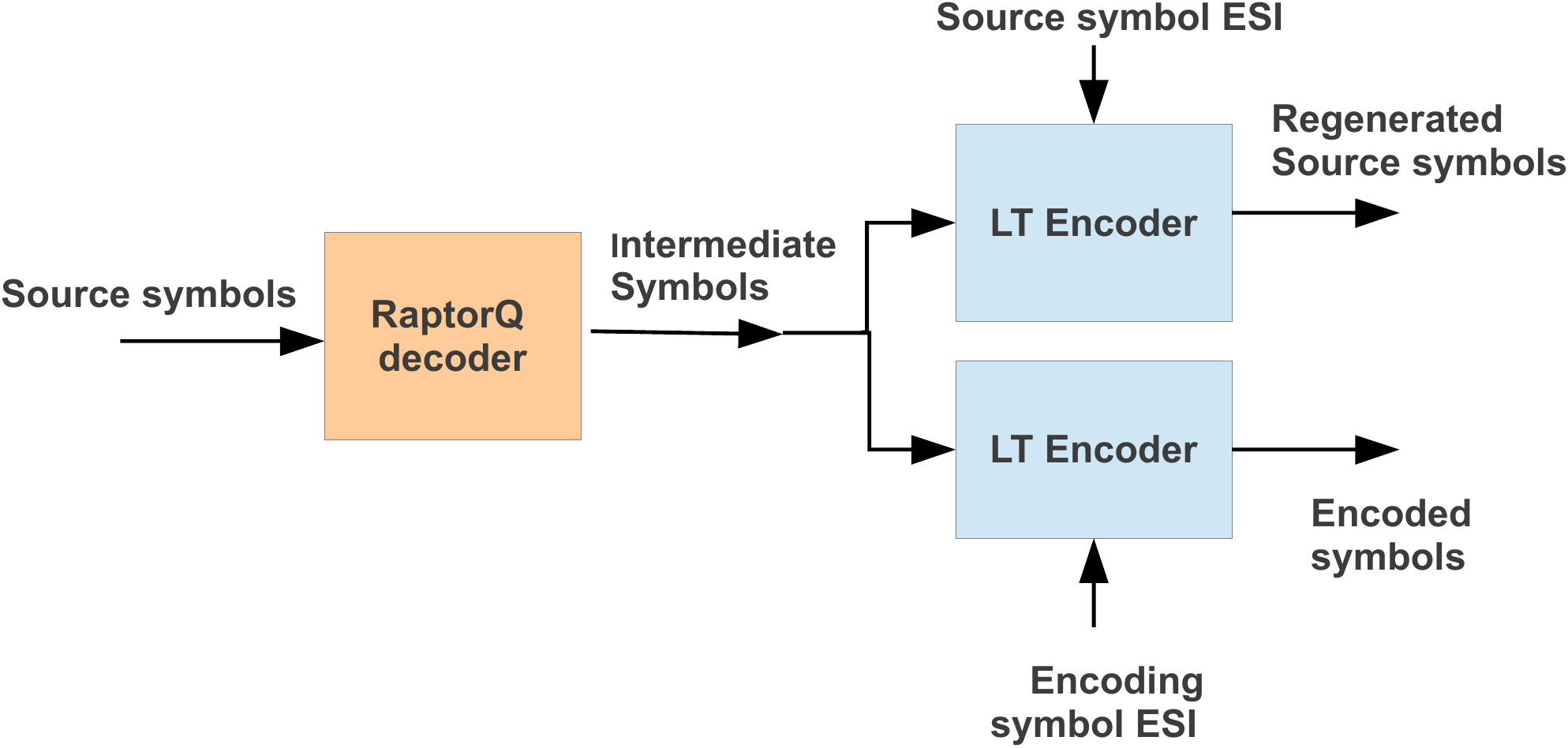}
       \label{fig:encoder_bd}
        \includegraphics[bb=0 -100 559 38, width=\columnwidth]{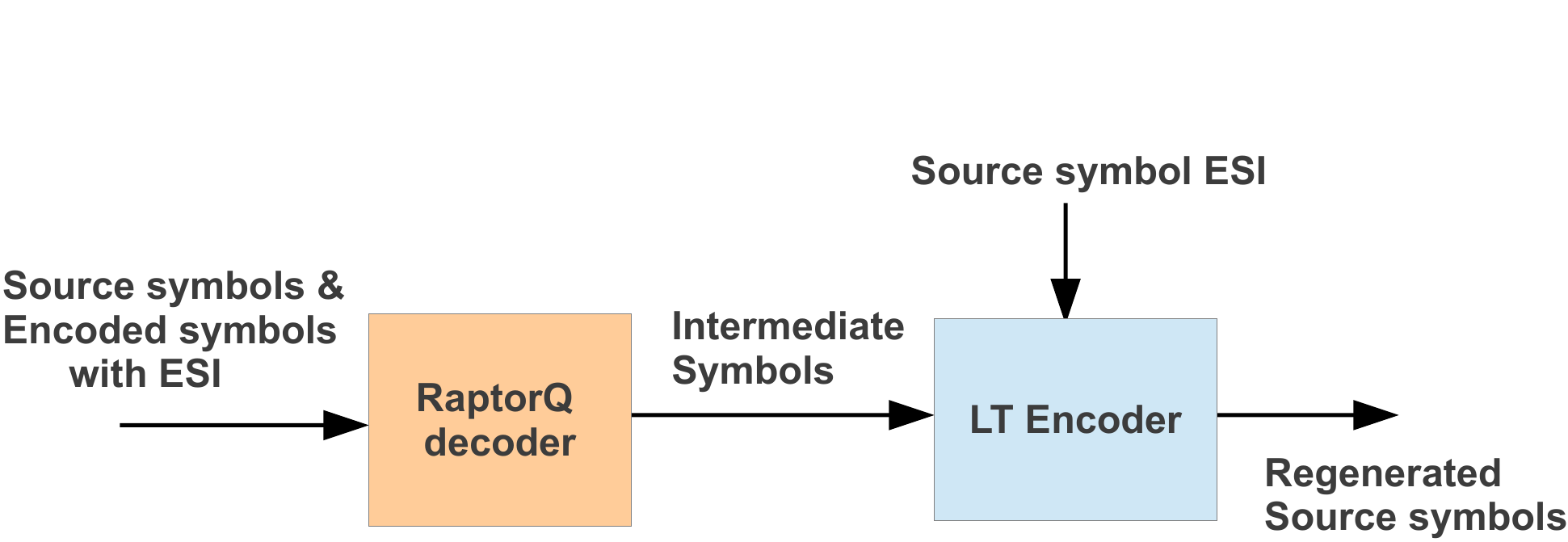}
        \label{fig:decoder_bd}
    \caption{Block diagrams showing RaptorQ encoding (left) and decoding (right) schemes}
    \label{fig:encoder_decoder_bd}
\end{figure*}

We propose a new UEP scheme called \emph{Priority Based Precode Ratio (PBPR)} 
for  standard systematic RaptorQ and Raptor codes. PBPR assumes that all symbols in a source block belong to the same importance class.
The UEP is achieved by changing the number of precode symbols \emph{for the entire block} based on the priority of source symbols in
that block.  The number of decoding overhead symbols  is the same across all source blocks---even those belonging to different importance classes.
A salient feature of our scheme is that it seamlessly integrates into the standard design of RaptorQ and Raptor codes. 

Our work was motivated by the important application area of MPEG transmission \cite{3gpp.26.234}.
We focus our experimental design on source block sizes that are
most applicable to MPEG transmission. We show that for such block sizes, PBPR
achieves \emph{consistent} improvements in error correction capability for the
high priority symbols compared to lower priority symbols across the entire
range of erasure probabilities.

The  contributions of our  work are outlined below.  
\begin{enumerate}
\item We construct a new UEP scheme for standard systematic RaptorQ and Raptor codes based on priority-based precoding.
\item Our UEP scheme demonstrates consistent and significant improvement in
error correction capability for high priority symbols compared to lower
priority symbols.  
\item We empirically characterize the
improvement in error correction performance as a result of increasing the
number of precode symbols in RaptorQ codes. \end{enumerate}

\section{Background} 
RaptorQ codes are systematic codes whose failure probability
curve is similar to that of random binary fountain codes. The systematic property
of Raptor family of codes is achieved by creating a set of \textit{intermediate symbols},
which are generated by "decoding" the original source symbols at encoder. 
These intermediate symbols are encoded to generate the original source symbols
and repair symbols. The intermediate symbols are regenerated at the decoder from the set of
received encoding symbols and the original source symbols are calculated from 
these intermediate symbols. Each symbol (both repair and source symbol) is identified 
by a unique 24 bit integer identifier known as \textit{encoding symbol identifier} (ESI),
which is carried from  encoder to decoder as part of the symbol overhead.
The ESI is used for generating the pseudo-random distribution which is applied 
to calculate the repair symbol degree and to derive the set of source symbols
to generate the symbol. The block diagrams showing the encoding and decoding
are captured in Fig.\ref{fig:encoder_decoder_bd}.\\

\begin{figure}[!b]
\centering
\includegraphics[scale=0.3]{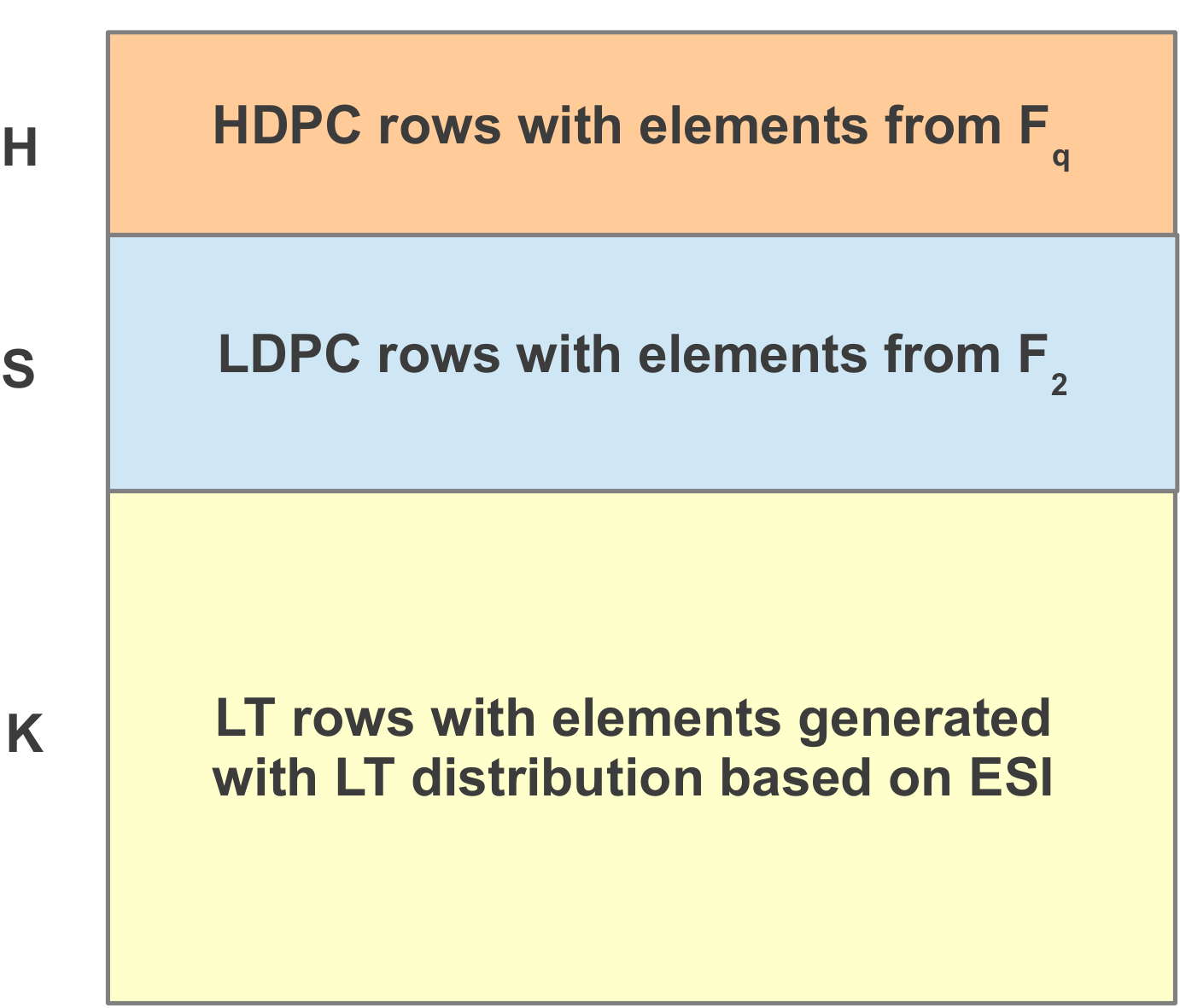}
\caption{RaptorQ generator matrix structure, where $H$ is the number of HDPC symbols, $S$ is number of LDPC symbols
        and $K$ is the number of LT symbols} \label{fig:raptorq_matrix}
\end{figure}

The precode of RaptorQ codes consists of two codes. An LDPC 
code, which is generated from a cyclically shifted circulant matrix and
another dense form of the LDPC code known as Half Density Parity-Check (HDPC)
code. The relationships of the intermediate symbols with the encoded symbols and 
constraint relations within the intermediate symbols are represented 
by a matrix, which is known as the \textit{generator matrix}. 
Since the precoding phase is combined with the LT coding in Raptor family of codes,
the generator matrix is created from three sub-matrices. The first set of rows correspond to HDPC 
codes, the next set correspond to LDPC code and rest of the rows correspond to LT codes
and generated from LT distribution. The schematic representation of generator matrix
is shown in Fig. \ref{fig:raptorq_matrix}.\\

Raptor and RaptorQ codes use a special type of decoding method known as
\textit{inactivation decoding} \cite{Shokr09}, which is a combination of belief propagation 
decoding and gaussian elimination. In practice, the decoding is done by simplification of generator matrix
which involves row and column reordering and row operations over $\mathbb{F}_q$. 
The overhead symbols (additional encoding symbols received by decoder) add 
more rows to the generator matrix, but these additional rows
subsequently get dropped after choosing the set of independent rows at the end of
first phase of decoding \cite{rfc5053}. \\

RaptorQ codes support larger block sizes, superior flexibility and better coding efficiency
compared to Raptor codes due to two major enhancements in the coding scheme. 
First enhancement is the usage of symbols from larger finite fields. 
The elements of HDPC rows in RaptorQ codes are chosen from $\mathbb{F}_{256}$ and the rest of 
the rows are from $\mathbb{F}_{2}$. Using a subset of symbols from $\mathbb{F}_{256}$
improves the failure probability curve, without causing large performance penalty \cite{Shokr09}.
Second enhancement is the change in generator matrix structure 
which introduces a variation to inactivation decoding known as \textit{permanent inactivation},
wherein a subset of columns in the LT part of the generator matrix are 
generated with uniform random distribution.\\

\section{Related work:UEP for Fountain \& Raptor codes}
A novel scheme of UEP for Raptor family of codes was first proposed by Rahnavard et. al. \cite{RahVF07}. 
It is a generic scheme that can be used for any Fountain code. This method was subsequently
improved to produce different variations, such as the one proposed in \cite{YuaA10}. 
The UEP is achieved by modifying the selection probability of the source symbols for 
check symbol generation in these schemes. The precode (LDPC encoding/decoding)
and LT coding are treated as two separate independent phases and UEP 
is applied either at LT coding (UEP-LT codes) or at precode (UEP-LDPC codes) or both.
The underlying idea is to select the source symbols 
from higher importance classes  with higher probability compared to 
those from lower importance classes for check symbol generation. Although such schemes are
very effective, they cannot be used with standard Raptor codes which are systematic. \\

Another method used for achieving UEP in case of fountain codes is
to divide the information symbols into multiple overlapping and expanding windows. The
symbols with higher priority are made part of more number of these windows
compared to symbols of lower priority, which increases the recovery probability
of symbols with higher priority. This method is applied to Raptor codes by 
partitioning the source symbols into multiple overlapping windows to achieve better error correction \cite{Sejdin09}.

UEP for systematic Raptor codes has been proposed recently with a new
design for Raptor codes \cite{Noh13}, which is different from standard design. The intermediate 
symbols are generated in a manner which is similar to the standard  scheme with a generator matrix 
and these symbols are encoded using separate LDPC and LT encoding steps to generate encoding symbols. 
The UEP is achieved by partitioning encoding matrices into different sub-matrices and 
changing the properties of these sub-matrices to group the symbols into different important classes. \\

\section{proposed UEP scheme: Priority Based Precode Ratio} 
We propose a new scheme of UEP for RaptorQ codes, which we call priority based precode ratio (PBPR). 
This scheme is designed with the assumption that all symbols in a 
source block (set of source symbols which participate in encoding and decoding as a unit) 
belong to the same importance class. The UEP is achieved by modifying the precode
properties of the RaptorQ codes. 
\subsection{Key Idea}
\begin{figure}[H]
\centering
\includegraphics[scale=0.3]{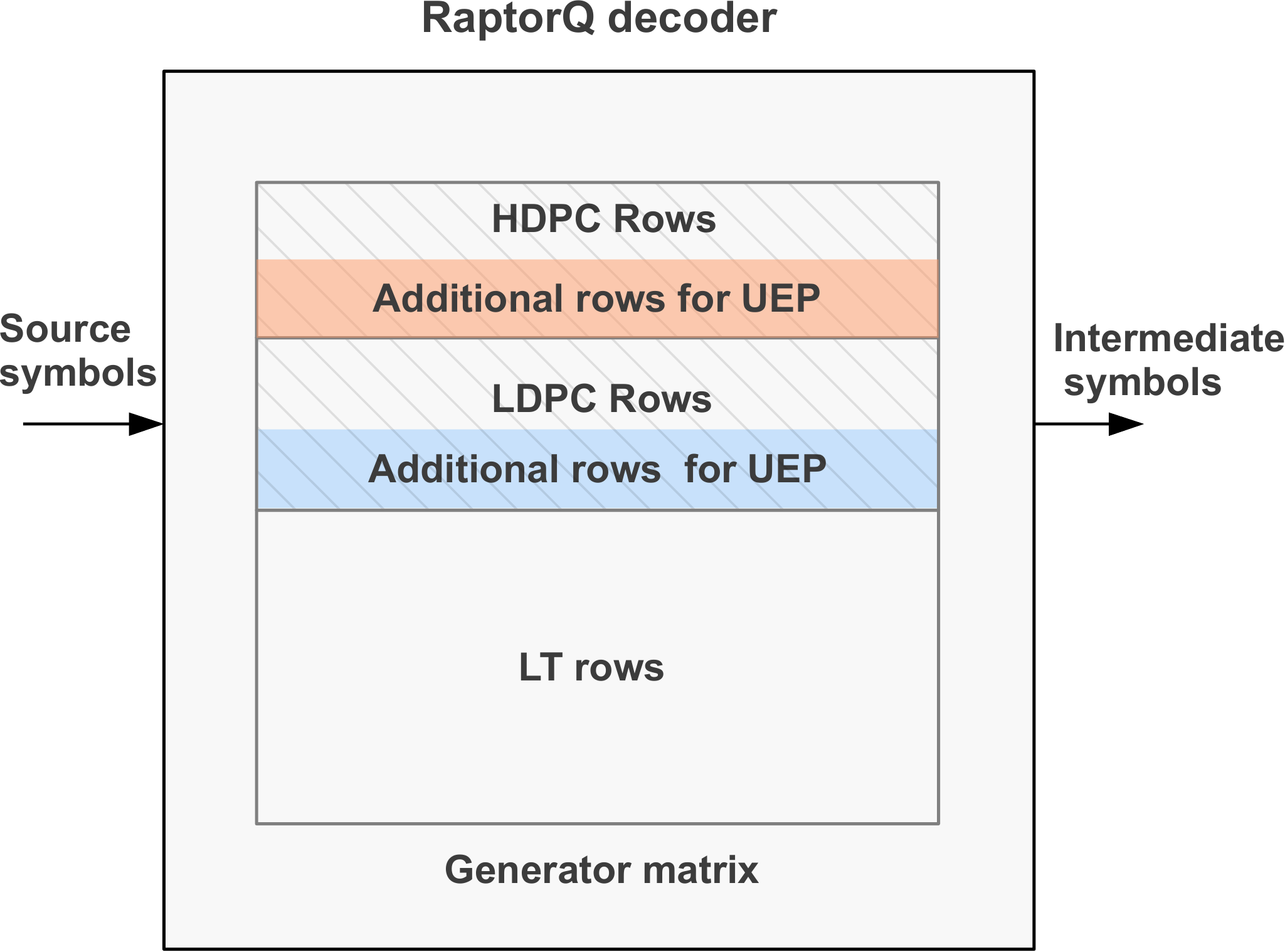}
\caption{RaptorQ decoder highlighting the changes done to achieve UEP by PBPR method. Number of precode rows are made 
        proportional to the priority of symbols in source block.} \label{fig:uep_scheme}
\end{figure}

We empirically demonstrate that increasing the number of precode constraint symbols of RaptorQ codes 
increases the error correction capability and make use of this property to design UEP. 
The UEP is achieved by making source block with symbols from higher importance class 
to have more number of precode symbols compared to the source block with symbols 
from lower importance class. In our preliminary work, we modified both HDPC and LDPC 
symbol numbers for designing the UEP scheme. The precode constraints of Raptor codes 
are designed such that the symbols which are not recovered in LT decoding process 
are recovered by the precode with very high probability. Increasing the number of precode 
symbols increases the probability of recovery for those symbols which were not decoded by LT 
decoding. The Fig. \ref{fig:uep_scheme} diagrammatically shows the changes introduced
in the RaptorQ decoder for the design of UEP scheme.\\

\subsection {Theoretical basis}
The Raptor coding scheme relies on generation of intermediate symbols
from source symbols (or from encoded symbols, in case of decoder). Once the intermediate
symbols are generated, original source symbols can be regenerated based on the ESI 
of the source symbols. Generation of the intermediate symbols from encoded 
symbols is analogous to solving a set of simultaneous equations, wherein the intermediate 
symbols act as the independent variables which are being solved. Any method of solving 
linear system of equations can be applied to decode the intermediate symbols. 
The linear system equations can be solved only when the associated coefficient matrix is of full rank. 
This property makes the error correction probability of Raptor codes to be dependant
on the rank profile of the generator matrix. The decoding of Raptor codes is
successful if the decoder matrix is full rank and all the intermediate symbols are
successfully recovered. The decoding completely fails if the matrix is full rank
and none of the symbols are recovered. Therefore, all the symbols in source
block which are part of decoding step have equal probability of being decoded and thus have
uniform error protection.\\

We analyse the effect of increasing the HDPC precode symbols on the rank profile
of generator matrix and show that the probability of matrix being full rank
increases with increase in precode rows. 
In addition to improvement in the rank profile of the generator matrix,
increase in the precode symbols improves the error correction capability of the 
RaptorQ codes in general. The precode rows of generator matrix define a 
set of LDPC constraint relations within the intermediate symbols. 
Increase in the precode rows increases the number of these constraints,
which increases the probability of intermediate symbol recovery in decoding.\\

\begin{figure}[h]
\centering
\includegraphics[scale=0.3]{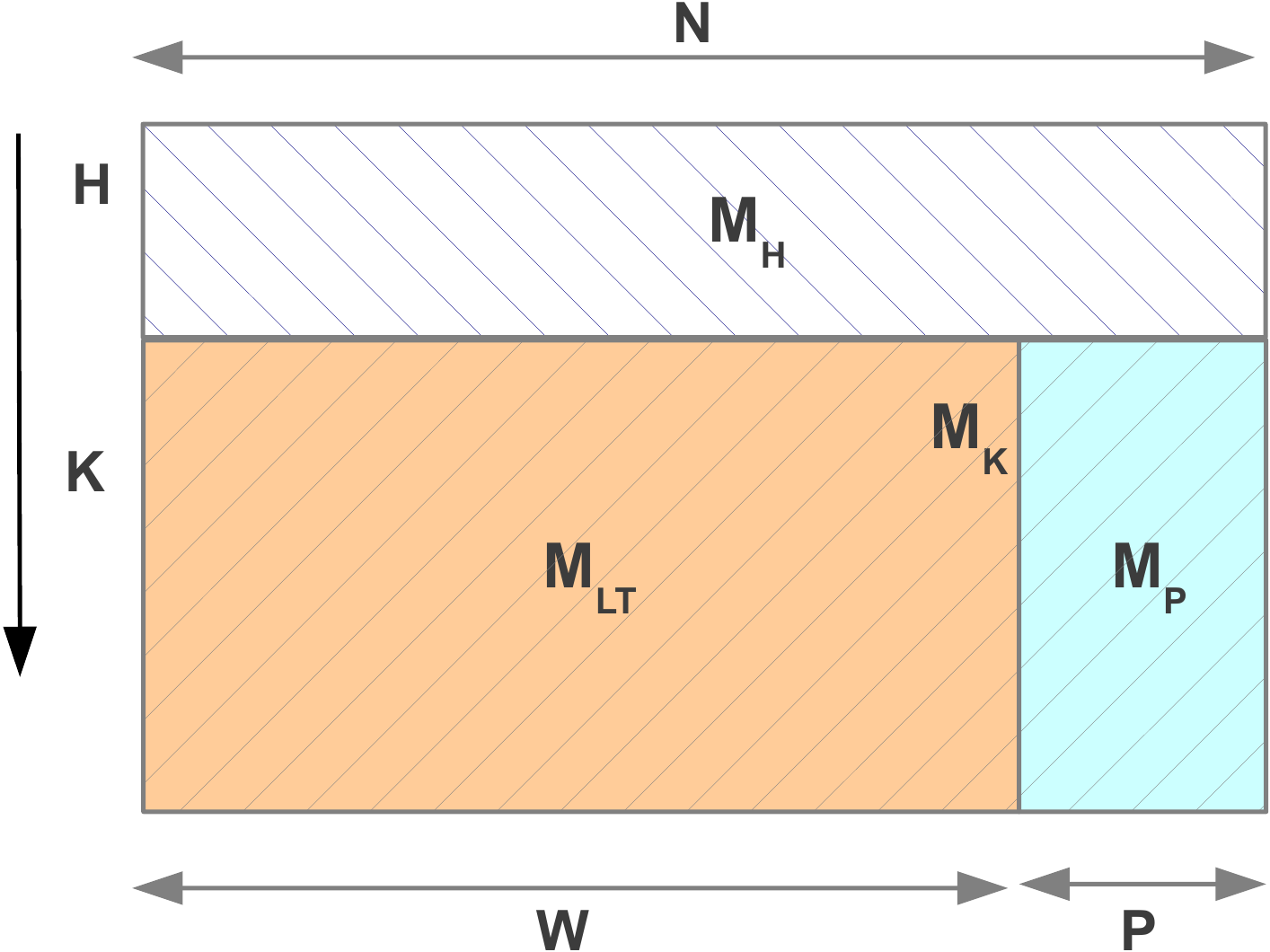}
\caption{Matrix $A$ considered for rank profile analysis based on the increase in HDPC rows} \label{fig:rank_analysis}
\end{figure}

To analyse the effect of increasing the HDPC symbols on the rank profile of the generator matrix,
let us consider a RaptorQ generator matrix $A$ with $K$ LT rows, $H$ HDPC rows and $N$ columns, 
as shown in Fig.\ref{fig:rank_analysis} .
For this analysis, we only reproduce the aspects of the derivation that we require, 
and refer the reader to \cite{Shokr09} for details of the derivation. 
Let the probability of LT sub-matrix ($K$ rows of the matrix) having full rank 
(rank of $K$) be $p_{K}$. This probability is same as that of standard RaptorQ codes
since there is no change in this sub-matrix. Rest of the rows (HDPC precode rows) 
of the matrix have their elements chosen such that their properties are similar to the random
binary matrix.  The probability of such a matrix (matrix $M_H$ in Fig. \ref{fig:rank_analysis}) 
of precode rows having full rank is given by \cite{Coop00},\cite{BlomKW97}

\begin{equation} 
p_r = \prod_{i=1}^{H}(1-\frac{1}{q^i}),
\end{equation}

where $q$ is the order of finite field, which is $256$ in this case.

Let the number of LT columns (the columns corresponding to LT distribution)
be $W$ and number of permanent inactivation (PI) columns be $P$ in the combined
LT matrix $M_{K}$. The PI columns are binary and also appear as to be chosen uniformly at random.  
In the matrix $A$, $K$ rows are considered as LT rows and $H$ rows are the additional 
rows due to precode.  The PI columns are the additional
columns outside the LT distribution of the combined matrix $M_{K}$. 
Since it is an LT decoder matrix, the additional overhead on LT
symbol rows due to the difference in number of rows and LT columns is described as

\begin{equation*} 
K - W = N - W - H = P - H 
\end{equation*} 
These are the additional rows which are introduced because of
PI symbols.  
Let $p_W$ denote the probability that the rank of the matrix $M_{LT}$
is $W$. Then the rank probability of the combined matrix $M_{K}$ is given by 
\begin{equation*} 
p_{K}= p_{W}\prod_{i=1}^{P-H-1}(1-2^{i-P}). 
\end{equation*} 

The probability $p_K$ may be approximated by 
\begin{equation*} 
p_{K}= p_{W}(1-2^{-H}) 
\end{equation*}

To take the additional rows (which are from $\mathbb{F}_q$) into account, the combined
probability of matrix $A$ having full rank can be written as 
\begin{equation} 
p_{N} = p_{K}\prod_{i=1}^H (1- \frac{1}{q^i} ) =
    p_{W}(1-2^{-H})\prod_{i=1}^H (1- \frac{1}{q^i} ). \label{eq:pN}
\end{equation}
See \cite{Shokr09} for further details of the derivation above.

As can be seen from \eqref{eq:pN}, the value of $p_N$ increases as $H$ is increased, 
since the value of $p_K$ is constant for our analysis.  Since $q$ is much larger
than 2, adding terms to the second part of the equation grows much smaller than
$(1-2^{-H} )$ term. 
Therefore, the rank probability of matrix increases as the number of $H$ rows is
increased. 

\subsection{Increase in decoding time}
Increasing the precode constraints increases the number of intermediate symbols and
in effect increases the number of dense rows in the RaptorQ generator matrix. 
These dense rows are more probable to increase the number of rows of the matrix
which is used for gaussian elimination phase of decoding. Addition of rows to gaussian elimination 
phase increases the computation cost, since the  complexity of gaussian elimination is quadratic ($O(n^2)$) 
compared to the linear time complexity of belief propagation. However, since the increase 
in precode size is relatively small compared to the number of LT symbols, the decoding 
time is comparable with the standard RaptorQ implementation and there is no significant increase
in decoding cost.

\section{Experimental results}
The PBPR method was simulated for RaptorQ codes considering two importance
classes which are commonly known as \textit{more important bits} (MIBs) and \textit{less important bits} (LIBs).
The simulation was done for different source block sizes and
different erasure probabilities. 
The block sizes we considered range from 53 bytes up to 213 bytes. As stated earlier, these
block sizes model actual usage in MPEG transmission \cite{3gpp.26.234}. 

We followed the method proposed in \cite{rfc6330} for simulating 
the encoder and decoder for RaptorQ codes, including the finite field arithmetic. 
The set of source symbols with block size was passed as input to the encoder. 
The decoder logic implemented in the encoder side generated intermediate symbols
and these intermediate symbols were verified by regenerating source symbols by encoding them.
The encoder generated the encoder symbols with ESI values which are greater than those which are
used for source symbols. We used ESI values ranging from 0 to $2^{24}-1$ for the encoding symbols 
so that sufficient randomness is introduced in the generation of LT distribution, which uses
ESI value as the basis.\\
The erasure channel was simulated by dropping (erasing) the source symbols and encoded symbols uniformly at random
with given erasure probability. Both source symbols and encoded symbols underwent same erasure 
mechanism to make sure that the behavior is comparable to the real-world behavior. We considered 
erasure values ranging from 0.1 up to 0.9 for our experiments.
The decoder received the mixture of source and encoded symbols which survived the channel erasure
and these symbols are used to generate the LT part of the decoder matrix based on ESI values. 
The precode part of the matrix was generated based on the rules proposed in \cite{rfc6330}.
Most of the simulations were done for the scenarios considering zero overhead symbols at the decoder,
except for one case, in which we considered one overhead symbol. Our experiments were done with the source
block sizes ranging from 55 bytes up to 213 bytes. 
In our preliminary investigations, we modified the precode sizes for higher importance classes such that
the error correction capability is better than lower importance classes but the cost in terms of performance 
is not significant. Table. 1 captures the values we used for the higher importance class
in comparison with lower importance class.\\

\begin{table}[h]
\centering
\caption{ Comparison of precode symbol sizes for different source block sizes in PBPR method }
\begin{tabular}{ccccc}
\toprule
K (LT Rows) & \multicolumn{2}{c}{S (LDPC rows)}  & \multicolumn{2}{c}{H (HDPC rows)} \\
                  &    LIB    &  MIB      &   LIB      &    MIB    \\ 
\midrule
             55   &    13     &  17       &   10       &    12     \\ 
             101  &    17     &  23       &   10       &    12     \\ 
             213  &    23     &  27       &   10       &    12     \\ 
\bottomrule
\end{tabular}
\label{tab:table1}
\end{table}

Note: Each experiment was run  $10^6$ times, for 
 erasure probabilities ranging from $0.1$ to $0.9$ in steps of $0.1$.
Recall that $K$ denotes the source block size. 
\subsection{Experiment 1: UEP for  $K = 55$ bytes, zero overhead symbols }
{\noindent\textit{Objective.}} To conduct the error probability comparison experiment with source block size of 55 bytes, which was chosen 
from the standard \cite[Table. 2]{rfc6330}. \\
{\noindent\textit{Methodology.}}  For the LIB source block, the precode sizes were
same as suggested by standard, i.e. $H=10$ and $L=13$. For the MIB symbol source block
we increased the HDPC precode size to $H=12$ and LDPC precode size to $S=17$. \\
{\noindent\textit{Results.}} The results are captured in Fig. \ref{fig:fig_k_55} showing both LIB and MIB
error correction probability vs. different erasure probabilities. \\
{\noindent\textit{Discussion.}} The results show that the MIB error performance is improved approximately
up to 1.4 times compared to the LIB error performance.
\subsection{Experiment 2: UEP for  $K = 101$ bytes, zero overhead symbols }
{\noindent\textit{Objective.}} To conduct the error probability comparison experiment with source block size of 101 bytes, which was chosen 
from the standard \cite[Table. 2]{rfc6330}. \\
{\noindent\textit{Methodology.}} For the LIB source block, the precode sizes were
same as suggested by standard, i.e. $H=10$ and $L=17$. For the MIB symbol source block
we increased the HDPC precode size to $H=12$ and LDPC precode size to $S=23$.  \\
 {\noindent\textit{Results.}} The results are captured in Fig. \ref{fig:fig_k_101} showing both LIB and MIB
error correction probability vs. different erasure probabilities. \\
{\noindent\textit{Discussion.}} The results show that the MIB error performance is improved approximately
up to 1.4 times compared to the LIB error performance.
\subsection{Experiment 3: UEP for  $K = 213$ bytes, zero overhead symbols }
{\noindent\textit{Objective.}} To conduct the error probability comparison experiment with source block size of 55 bytes, which was chosen 
from the standard \cite[Table. 2]{rfc6330}. \\
{\noindent\textit{Methodology.}} For the LIB source block, the precode sizes were
same as suggested by standard, i.e. $H=10$ and $L=23$. For the MIB symbol source block
we increased the HDPC precode size to $H=12$ and LDPC precode size to $S=27$.  \\
{\noindent\textit{Results.}} The results are captured in Fig. \ref{fig:fig_k_213} showing both LIB and MIB
error correction probability vs. different erasure probabilities. \\
{\noindent\textit{Discussion.}} The results show that the MIB error performance is improved approximately
up to 1.6 times compared to the LIB error performance.
\subsection{Experiment 4: UEP for  $K = 101$ bytes, one overhead symbol}
\begin{figure}[H]
\centering
    \begin{subfigure}{\columnwidth} 
        \centering
        \includegraphics[scale=0.35]{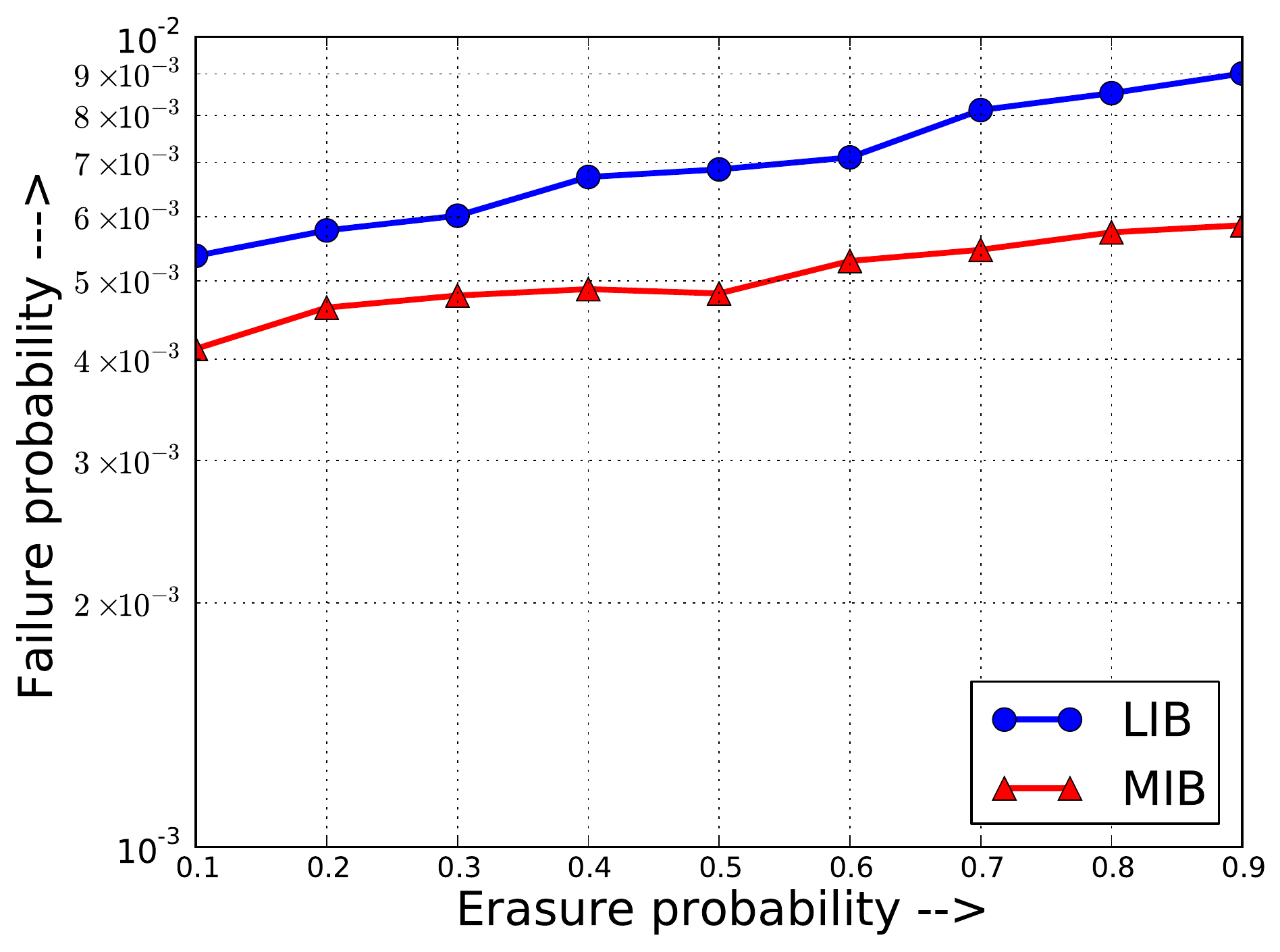}
        \caption{Source block size of 55 bytes.}
        \vspace{12pt}
        \label{fig:fig_k_55}
    \end{subfigure}
    \begin{subfigure}{\columnwidth} 
        \centering
        \includegraphics[scale=0.35]{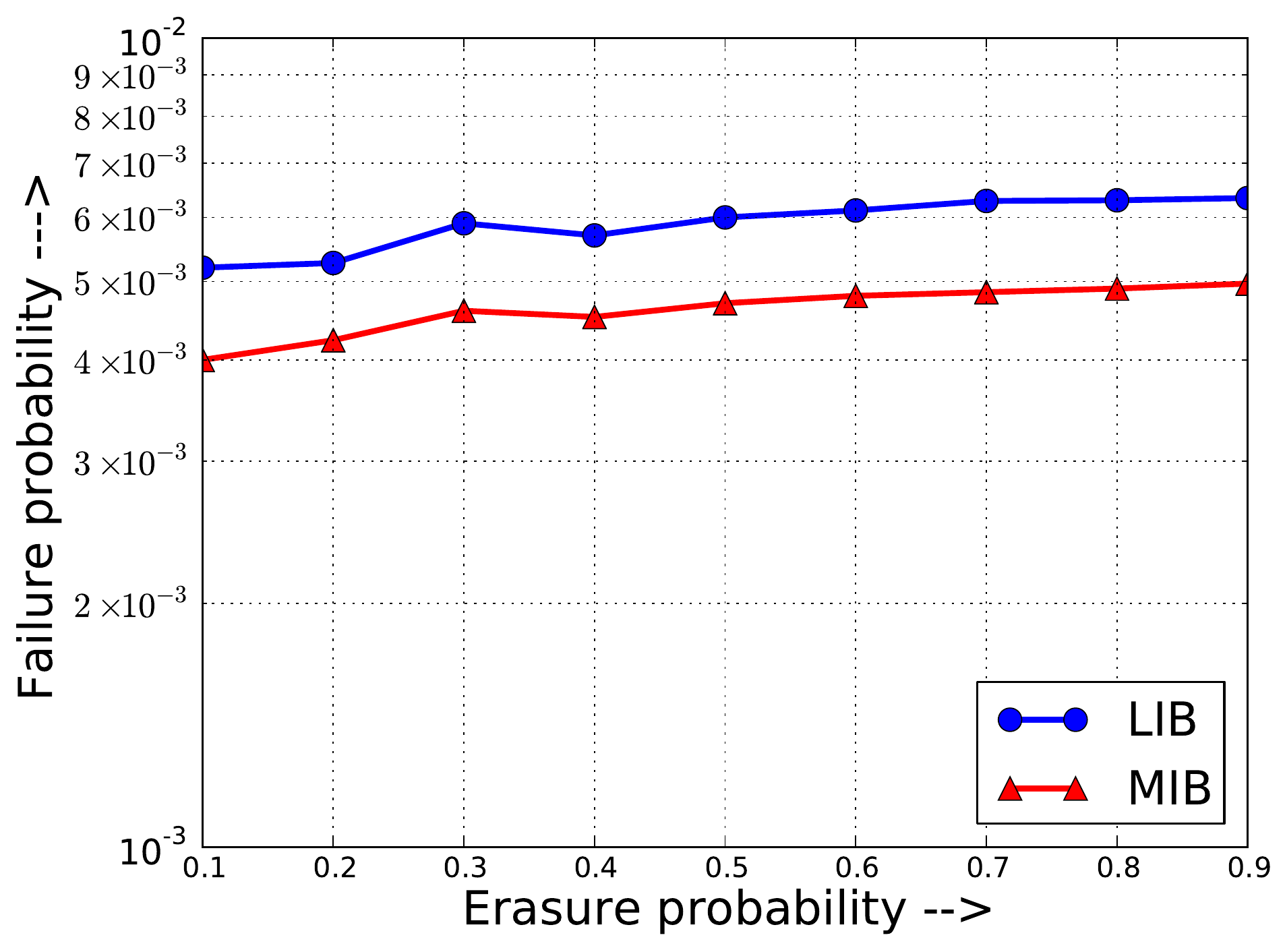}
        \caption{Source block size of 101 bytes.}
        \vspace{12pt}
        \label{fig:fig_k_101}
    \end{subfigure}
    \begin{subfigure}{\columnwidth} 
        \centering
        \includegraphics[scale=0.35]{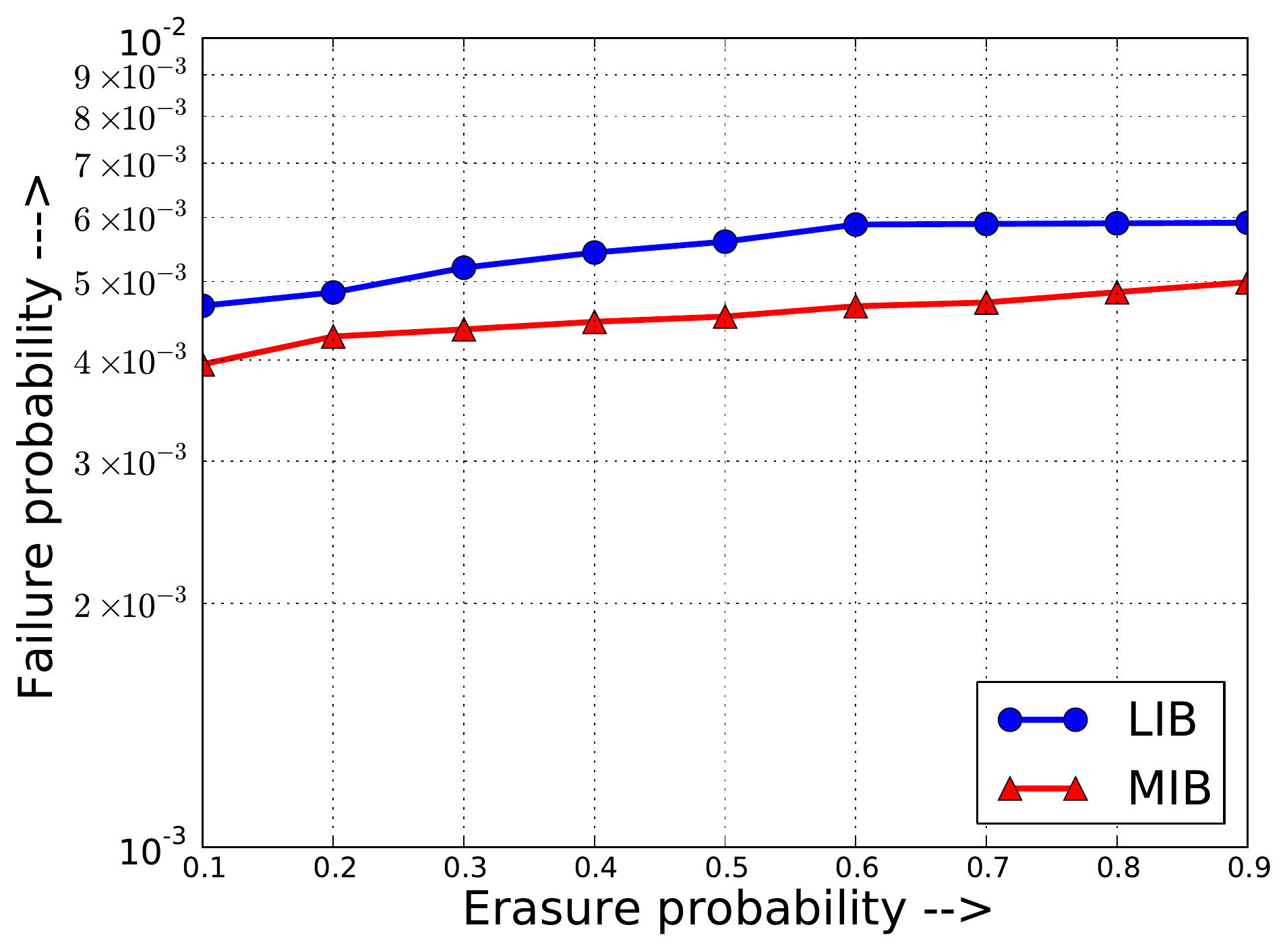}
        \caption{Source block size of 213 bytes.}
        \vspace{12pt}
        \label{fig:fig_k_213}
    \end{subfigure}
    \begin{subfigure}{\columnwidth} 
        \centering
        \includegraphics[scale=0.35]{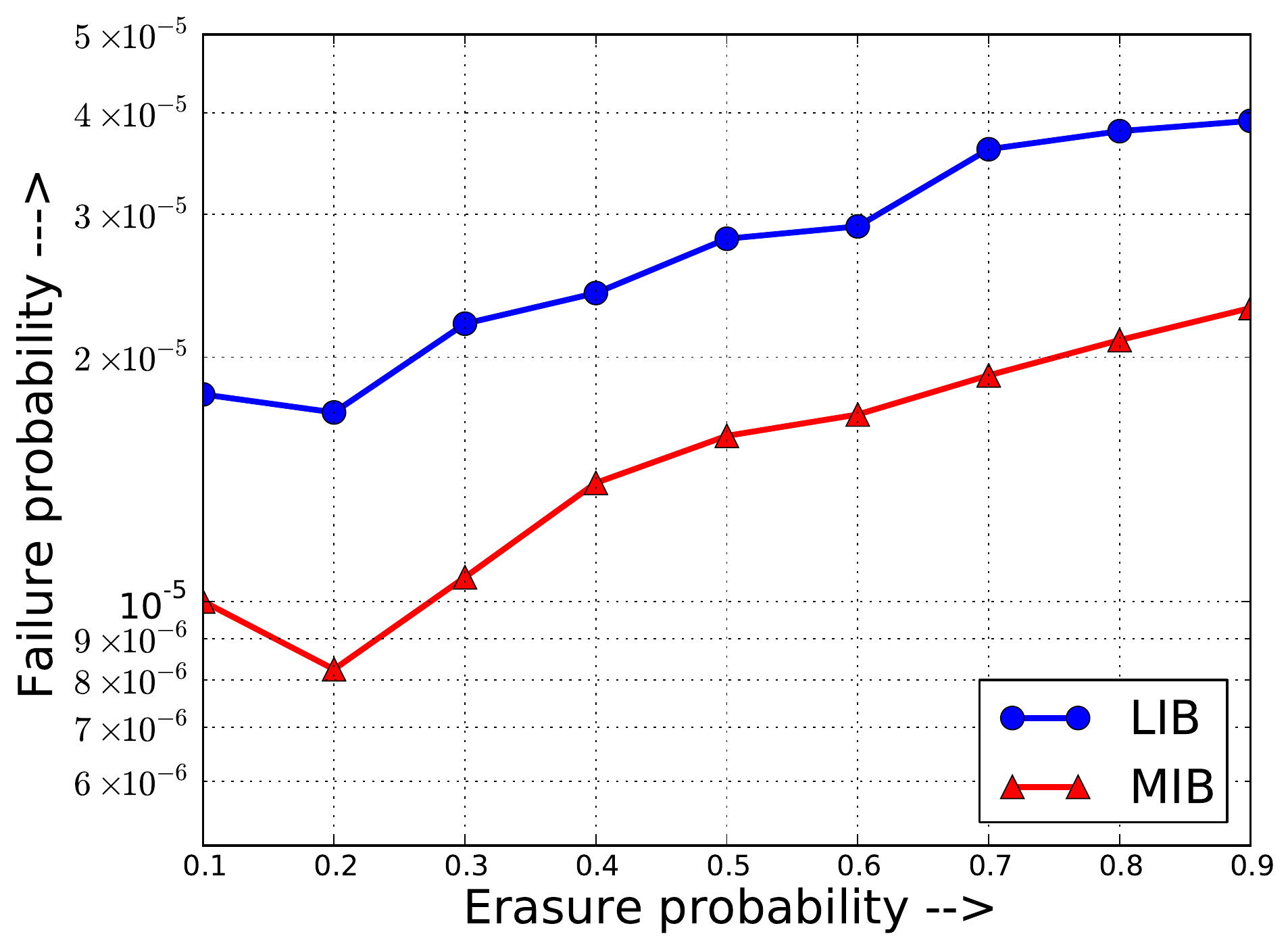}
        \caption{Source block size of 101 bytes with 1 overhead symbol.}
        \label{fig:fig_k_101_e_1}
    \end{subfigure}
\caption{Comparison of MIB and LIB performance for different source block sizes. The failure probability of MIB and LIB symbols is shown for 
different erasure probabilities from $0.1$ to $0.9$ in steps of $0.1$.}
\label{fig:sim_result}
\end{figure}

{\noindent\textit{Objective.}} To conduct experiments with source block size of 101 bytes 
and 1 overhead symbol. \\
{\noindent\textit{Methodology.}} For the LIB source block, the precode sizes were
same as suggested by standard, i.e. $H=10$ and $L=17$. For the MIB symbol source block
we increased the HDPC precode size to $H=12$ and LDPC precode size to $S=23$. The 
encoding symbols used for decoding was $K+1$ (102) in this case.  \\
{\noindent\textit{Results.}} The results are captured in Fig. \ref{fig:fig_k_101_e_1} showing both LIB and MIB
error correction probability vs. different erasure probabilities. \\
{\noindent\textit{Discussion.}} The results show that the MIB error performance is improved approximately
up to 2 times of the LIB error performance.
\begin{table}[hb] 
\centering
\caption{ Percentage increase  in encoding-decoding time with PBPR for different source block sizes}
\begin{tabular}{cc}
\toprule
 Source Block size &  Increase in Decoding Time \\
(bytes) &  (percentage) \\
\midrule
 55  & 18     \\ 
 101 &  13.35 \\
 213 &  7.5   \\
\bottomrule
\end{tabular}
\label{tab:table2}
\end{table}

\subsection{Experiment 5: Increase in  encoding-decoding time with PBPR}
{\noindent\textit{Objective.}} To conduct experiments to measure the increase in encoding and decoding time 
because of increasing the precode size. \\
{\noindent\textit{Methodology.}} The simulations were done for the LIB and MIB precode values
specified in Table \ref{tab:table1}. The time taken for simulation was measured by averaging over 
5000 runs. The percentage increase in the MIB encoding/decoding time for different source
block sizes are captured. \\
{\noindent\textit{Results.}} The results in the form of percentage increase in time are captured in Table \ref{tab:table2}.\\
{\noindent\textit{Discussion.}} The table shows that there is some increase in the performance, which is up to
18\% increase in the time taken by MIB symbol encoding and decoding. This performance penalty is
proportional to the ratio of increase in precode symbols to the LT symbols. 
This penalty seems to be decreasing as the ratio is decreased.

\section{Conclusion} 
We propose a new UEP scheme called PBPR for RaptorQ and Raptor codes. PBPR is 
designed with the assumption that all symbols in a source block belong to same importance class. 
 UEP is achieved by changing the number of precode symbols based on the priority of symbols in source block. 
We empirically show that increasing the number of precode symbols increases the error correction capability 
of RaptorQ codes. 
Our experimental results confirm a consistent and significant increase in error correction capability of higher importance symbols compared to lower importance ones.  We also show that the increase in encoding and decoding time using PBPR is insignificant compared to standard implementations.

In the longer version of the paper, we will report on a significantly expanded experimental design, and also include a complete theoretical treatment for the improvement of error 
correction capability witnessed by using PBPR.

\section*{Acknowledgment}
Authors wish to thank Nazanin Rahnavard for graciously providing 
helpful pointers to literature.

\bibliographystyle{IEEEtran}
\bibliography{references}

\end{document}